\documentclass{article}
\pdfoutput=1
\usepackage{arxiv}
\usepackage{lineno,hyperref}
\usepackage{booktabs} 

\usepackage{amssymb}
\usepackage{amsthm}
\usepackage{amsmath}

\usepackage{hyperref}

\usepackage{algorithm}
\usepackage{algpseudocode}

\usepackage{tikz}
\usetikzlibrary{plotmarks}
\usepackage{pgfplots}
\usepackage{svg}
\usetikzlibrary{positioning,shapes,arrows,shadows,patterns}

\usepackage{multirow}
\usepackage{colortbl}

\usepackage{amsmath,amsfonts}
\usepackage{array}
\usepackage[caption=false,font=normalsize,labelfont=sf,textfont=sf]{subfig}
\usepackage{textcomp}
\usepackage{stfloats}
\usepackage{url}
\usepackage{verbatim}
\usepackage{graphicx}

\title{Predicting Music Track Popularity by Convolutional Neural Networks on Spotify Features and Spectrogram of Audio Waveform}

\author{Navid Falah \\
  Department of Mathematics and Computer Science \\
  Amirkabir University of Technology\\ (Tehran Polytechnic)\\
  Iran\\
  \texttt{navid.falah@aut.ac.ir}\\
   \And
Behnam Yousefimehr \\
  Department of Mathematics and Computer Science \\
  Amirkabir University of Technology\\ (Tehran Polytechnic)\\
  Iran\\
  \texttt{behnam.y2010@aut.ac.ir}\\
   \And
  Mehdi Ghatee\footnote{The corresponding author} \\
  Department of Mathematics and Computer Science \\
  Amirkabir University of Technology\\ (Tehran Polytechnic)\\
  Iran\\
  \texttt{ghatee@aut.ac.ir}\\
}

\begin{document}
\maketitle
\begin{abstract}
In the digital streaming landscape, it's becoming increasingly challenging for artists and industry experts to predict the success of music tracks. This study introduces a pioneering methodology that uses Convolutional Neural Networks (CNNs) and Spotify data analysis to forecast the popularity of music tracks. Our approach takes advantage of Spotify's wide range of features, including acoustic attributes based on the spectrogram of audio waveform, metadata, and user engagement metrics, to capture the complex patterns and relationships that influence a track's popularity. Using a large dataset covering various genres and demographics, our CNN-based model shows impressive effectiveness in predicting the popularity of music tracks. Additionally, we've conducted extensive experiments to assess the strength and adaptability of our model across different musical styles and time periods, with promising results yielding a 97\% F1 score. Our study not only offers valuable insights into the dynamic landscape of digital music consumption but also provides the music industry with advanced predictive tools for assessing and predicting the success of music tracks.

\end{abstract}

\keywords{Music Track Popularity \and Deep Learning \and Feature Extraction \and Spotify Features \and Spectrogram of Audio Waveform}

\section{Introduction}

In the rapidly evolving digital music landscape, platforms such as Spotify have revolutionized how music is consumed, offering unparalleled insights into listeners' preferences\cite{anderson2020algorithmic,hodgson2021spotify}. This transformation presents an opportunity to harness vast datasets for analyzing trends and predicting song popularity, a pursuit that is increasingly crucial for artists and industry stakeholders aiming for market prominence\cite{pham2016predicting,lee2018music,zhang2024unveiling}. Our research leverages Spotify's extensive dataset, focusing on the top 100 artists of February 2023, to unveil patterns of musical popularity within the United States. By employing a Django-developed backend for efficient data extraction and the advanced analytics afforded by Convolutional Neural Network (CNN) models, this study transcends conventional methods, providing a nuanced understanding of song popularity driven by a comprehensive array of audio features.

Streaming services invest significantly in song suggestion algorithms, highlighting the importance of predictive analytics in music consumption\cite{o2020beyond}. Unlike previous studies that primarily used classification models, our approach utilizes regression to predict song popularity based on metrics calculated by Spotify. Emphasizing the multidisciplinary field of Hit Song Science (HSS)\cite{seufitelli2023hit}, our research intersects computer science with music theory, sociology of music, and cultural markets, analyzing music data across different modalities from various sources. Our main goal is to provide insights that could significantly benefit music companies by predicting a song's success before an album's launch, acknowledging the evolving nature of music tastes and the paramount importance of geographical and market contexts, especially focusing on the United States.\\
Our contribution lies in the comprehensive collection of rich data sets, coupled with our pioneering approach of simultaneously investigating regression and classification techniques to provide a holistic understanding of song popularity prediction.

\section{Related Work}
Machine learning encompasses the realm of artificial intelligence where algorithms are utilized to enable systems to learn from data and make predictions or decisions without explicit programming \cite{alpaydin2021machine}. When applied to predicting song popularity, machine learning algorithms scrutinize various features extracted from music tracks, encompassing audio characteristics, metadata, lyrics, and even metrics related to social media engagement. This analysis aims to forecast whether a song will attain popularity \cite{sebastian2024beyond}.

In the research cited, diverse machine learning algorithms have been investigated for song popularity prediction, including Random Forest \cite{breiman2001random}, logistic regression \cite{kleinbaum2002logistic}, Naïve Bayes \cite{an2017naive}, decision trees \cite{guggari2022music}, K-Nearest Neighbors \cite{kostrzewa2018classification}, and Support Vector Classifiers \cite{mandel2005song}. These algorithms dissect different facets of music tracks and associated metadata to unveil patterns indicative of popularity. For instance, audio features like tempo, energy, and danceability are examined to discern musical attributes that resonate with listeners, while metadata such as artist, genre, and release date furnish contextual insights. 

In the following, we will delve into pivotal studies in this domain and forecast the potential popularity of a song.

\cite{kaneria2021} implemented machine learning algorithms on features extracted from the Spotify database, focusing on supervised learning to classify tracks as hits or non-hits. Their methodology underscored the importance of data preprocessing and highlighted the potential of Random Forest, logistic regression, naïve Bayes, and decision tree algorithms in predicting song popularity.

Article \cite{yap2021} explored the integration of social media features from YouTube with Spotify's audio features to predict music popularity. Their study constructed a dataset of newly released tracks, defining music popularity through diverse metrics derived from Spotify's Top 200 daily chart. Their findings suggested that incorporating social media variables significantly enhances model performance, marking a novel approach in the Hit Song Science domain by jointly using audio features and social media data for predicting hit songs.

Article \cite{kamal2021} focused on song metadata and lyric analysis to predict song popularity, comparing several machine learning algorithms. Their research highlights the varied impact of metadata and lyric sentiment on the predictive accuracy, emphasizing the potential of combining these factors for enhanced prediction models.

Research \cite{pareek2022predicting} explored the use of Spotify's track metadata to predict song popularity, applying various machine learning algorithms such as Random Forest, K-Nearest Neighbors, and Linear Support Vector Classifiers. Their findings suggest that Random Forest achieved the highest accuracy, precision, recall, and F1-score among the classifiers tested, demonstrating the potential of using streaming data for predictive analytics in the music industry.

Yee \& Raheem \cite{yee2022predicting} used a random forest model on YouTube and Spotify's track metadata to predict the music popularity. They presented the accuracy, macro-precision, macro-recall and macro F1-scores  on released tracks from May to August 2021.

Gao \cite{gao2021catching} applied interpretable machine learning models to predict the music popularity  based on the audio features and artists. They asserted that the music industry can produce the music using controllable and predictable features that enourage the consumers.

The authores of \cite{martin2020multimodal} proposed an innovative multimodal end-to-end deep learning model (HitMusicNet) to predict popularity in music recordings. 

Saragih \cite{saragih2023predicting} employed regression and classification machine learning algorithms to thoroughly analyze audio features on Spotify that influence the popularity of streamed songs. This rigorous analysis assessed the significance of these features for accurate prediction.

Yin \cite{yin2023music} proposed a deep learning model for a music recommendation system. The model preprocesses the data, generates Mel spectrogram features, applies a convolutional neural network, and recommends music tracks based on user preferences.

Terroso-Saenz et al. \cite{terroso2023music} discovered a new way to analyze music mobility patterns using a directed-graph structure. The patterns correlate with migratory flows and cultural similarities among regions. This research could help record companies and artists with marketing and organizing events.

Furthermore, a study delving into the attribute-based approach in the realm of Hit Song Science (HSS) examined a diverse range of audio features from Spotify to understand their influence on song streams \cite{seufitelli2023hit}. This research underscored the complexity of predicting cultural product success, revealing that audio features had a modest but significant impact on song popularity.

Reisz et al. \cite{reisz2024quantifying} demonstrated how musical homophily predicts song popularity by analyzing last.fm data. The study explores how social connections influence song dissemination and popularity. Integrating musical homophily into machine learning models improves accuracy in predicting song popularity.

The authors of \cite{reddy2024cutting} developed a new music recommendation system using a Gravitational Search Optimized Recursive Neural Network (GS-RNN) to assess audio signal similarity for content-based recommendations.

Our current research builds upon these foundational studies, aiming to amalgamate and refine their methodologies by employing advanced neural processing techniques. By integrating a more extensive set of audio features with the latest in machine learning models, we aspire to develop a predictive model that not only forecasts song popularity with greater accuracy but also offers insights into the underlying patterns that define musical success in the digital age.

\section{Methods}
\begin{figure*}[t!]
\centerline{
\includegraphics[width=0.7\linewidth]{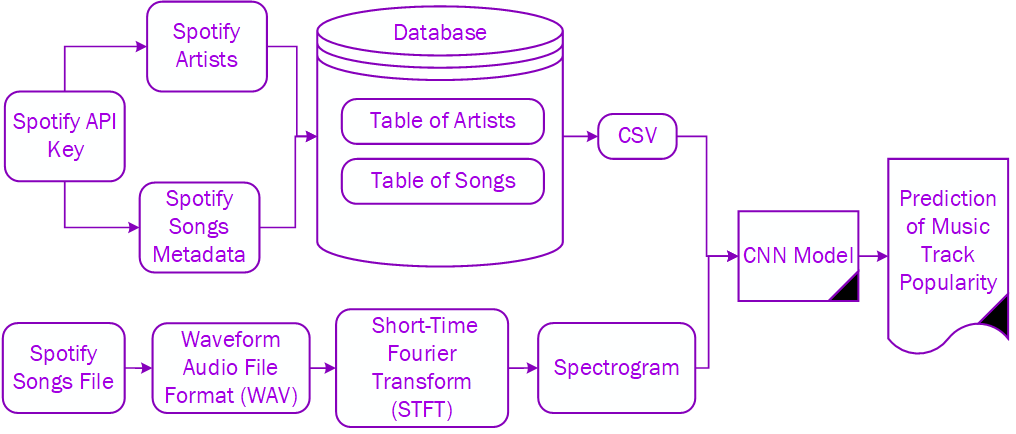}
}
\caption{Flow of Song Popularity Prediction Using Spotify Data and CNN.}
\label{fig:exampleImage}
\end{figure*}

 The methodology involved extracting data from the Spotify platform using the Spotify API key. This data encompassed details about artists and metadata related to songs. The information was organized into a database comprising two tables: one for artists and the other for songs. Figure~\ref{fig:exampleImage}  presents a schematic of the data collection and analysis process employed in the study.

A consolidated CSV file was created that included both artists and metadata for the songs. In addition to this, Spotify songs were converted to a WAV audio format for further analysis. These audio files were then transformed into spectrograms for feature extraction.

A convolutional neural network (CNN) model was utilized to process these features, aiming to predict the popularity of the tracks. The CNN model’s output was an estimation of the songs’ popularity, which is likely based on patterns learned from the spectrogram data and possibly other metadata. This approach allows the model to capture both the textual and acoustic characteristics that contribute to a song's appeal. By analyzing layers of data, the CNN can identify intricate patterns that correlate with higher popularity ratings, offering insights into what makes a song more likely to succeed in the market.

The model underwent rigorous training and validation phases, using a comprehensive dataset that included a wide variety of songs across different genres and epochs. This diversity in training data helped in enhancing the model's ability to generalize its predictions across unseen tracks, thereby increasing its accuracy and reliability. The end goal of the flowchart indicates the culmination of the process in which the model provides an estimated popularity metric for each song analyzed. Beyond merely predicting popularity, this metric could serve as a valuable tool for artists and producers to understand potential market reception and refine their creations accordingly. Furthermore, it opens pathways for future research into predictive analytics in music, potentially guiding the strategic release of songs to maximize audience engagement and success.

\begin{table*}[]  
\footnotesize
\centering
\caption{Summary of Music Features and Descriptions}
\label{tab:combined_music_features}
\begin{tabular}{lp{4.6cm}ccccc}
\hline
\textbf{Feature} & \textbf{Description} & \textbf{Mean} & \textbf{Std Dev} & \textbf{Min} & \textbf{Median} & \textbf{Max} \\
\hline
Popularity & Range of track popularity is (0-100). & 70.26 & 11.65 & 23 & 71 & 100 \\
Duration (ms) & Length of the track in milliseconds. & 208,487 & 48,666.6 & 40,690 & 206,120 & 613,027 \\
Acousticness & Confidence of track is acoustic (0-1). & 0.2276 & 0.2538 & 0.0000189 & 0.123 & 0.992 \\
Danceability & Suitability for dancing (0-1). & 0.6511 & 0.1418 & 0.177 & 0.664 & 0.965 \\
Energy & Intensity and activity measure (0-1). & 0.6477 & 0.1736 & 0.0297 & 0.667 & 0.995 \\
Instrumentalness & Likelihood of no vocals (0-1). & 0.0169 & 0.0939 & 0 & 0.00000129 & 0.96 \\
Liveness & Presence of an audience (0-1). & 0.1825 & 0.1393 & 0.0215 & 0.126 & 0.984 \\
Loudness & Loudness in decibels (dB). & -6.1834 & 2.3743 & -24.788 & -5.797 & -0.141 \\
Speechiness & Presence of spoken words (0-1). & 0.1005 & 0.0975 & 0.0232 & 0.0587 & 0.73 \\
Tempo & Beats per minute (BPM). & 121.18 & 28.48 & 59.635 & 119.994 & 208.038 \\
Time Signature & Beats per bar. & 3.9644 & 0.2664 & 1 & 4 & 5 \\
Valence & Musical positiveness (0-1). & 0.4832 & 0.2242 & 0.032 & 0.479 & 0.976 \\
Artist Popularity & Popularity of the artist (0-100). & 82.06 & 5.0974 & 69 & 82 & 100 \\
Artist Followers & Number of followers the artist has. & 28,574,700 & 23,531,300 & 494,115 & 23,141,900 & 115,073,000 \\
Release Year & Year the track was released. & 2016.85 & 7.6602 & 1970 & 2019 & 2024 \\
\hline
\end{tabular}
\end{table*}
\subsection{Data Collection}
Our investigation harnesses a dataset of approximately 2000 songs, inclusive of their associated artists, sourced from Spotify’s Web API. This dataset encapsulates a rich assortment of both audio features and metadata, each offering an in-depth perspective of the track, with a focus on the United States music industry. Significantly, this collection spans a broad spectrum of genres, reflecting the diverse musical landscape and ensuring the comprehensive analysis of patterns across different styles. To deepen the insight into artist trends and their impact on song popularity, we meticulously curated the dataset to include the top 5 hot tracks of each artist, recognized for their high streaming numbers or critical acclaim, alongside 15 additional tracks selected randomly. This strategy not only enriches our dataset with the artists' signature hits but also provides a broader understanding of their overall discography, totaling approximately 20 songs per artist across 100 artists.

To ensure robust model training and testing without data leakage, artist filtering was applied during dataset splitting. This method prevents any artist's tracks from appearing in both training and testing sets, thereby enhancing the generalizability of our predictive model.

We catalog a myriad of song characteristics, ranging from the mechanical—like duration and key—to more perceptual metrics such as danceability, energy, and Spotify’s own popularity scores. Additionally, our data acquisition process involves the utilization of Spotify's API to fetch detailed information on artists and songs. For song analysis, audio files are downloaded since their spectrograms are required for extracting various audio features. 

The dataset was meticulously curated, ensuring that only tracks with complete features were selected for an in-depth analysis. The metadata obtained through the API and the downloaded audio files are instrumental in our study, providing a comprehensive view of each song's attributes. 

The metadata attributes for chosen artists and song features are respectively outlined in Table \ref{tab:combined_music_features}, while Table \ref{tab:audio_features_castle} offers a glimpse into the audio features of "Castle" by Halsey. These tables provide insights into the elements scrutinized to comprehend a track's characteristics.

\begin{table}[H]
\centering
\caption{Audio Features for "Castle" by Halsey}
\label{tab:audio_features_castle}
\begin{tabular}{lp{10cm}l}
\hline
\textbf{Feature} & \textbf{Description} & \textbf{Example} \\
\hline
Acousticness & Confidence the track is acoustic (0-1). & 0.25 \\
Danceability & Suitability for dancing (0-1). & 0.626 \\
Duration\_ms & Duration in milliseconds. & 277623 \\
Energy & Intensity and activity measure (0-1). & 0.571 \\
Instrumentalness & Likelihood of no vocals (0-1). & 0.0 \\
Key & Key of the track (0-11). & 7 \\
Liveness & Presence of an audience (0-1). & 0.0946 \\
Loudness & Loudness in decibels (dB). & -7.461 \\
Mode & Modality (major-1, minor-0). & 0 \\
Speechiness & Presence of spoken words (0-1). & 0.0327 \\
Tempo & Tempo in BPM. & 129.959 \\
Time Signature & Beats per bar. & 4 \\
Valence & Musical positiveness (0-1). & 0.164 \\
\hline
\end{tabular}
\end{table}

\subsection{Data Preparation}
\label{sec:data_preparation}

During the data preparation phase, certain attributes were deemed non-essential for the predictive model and subsequently removed. These include the Spotify ID of the song, analysis URL, URI, and any additional links, which primarily serve as identifiers within the Spotify platform rather than features with predictive power for song popularity. Additionally, the 'key' attribute was also excluded from the analysis, as it showed negligible effect on the outcome of the predictions in preliminary assessments.

\subsection{Spectrogram Generation}

The transformation of audio waveforms into spectrograms is a critical step in feature extraction for predicting song popularity. We utilize the Mel Spectrogram, an alternative to the traditional Short-Time Fourier Transform (STFT), to convert time-domain audio signals into the frequency domain. This method adapts frequency resolution to align with human auditory sensitivity, which is more attuned to variations in lower frequencies. The Mel Spectrogram captures the signal’s frequency components over time, emphasizing perceptually significant features that are crucial for music analysis. This results in a two-dimensional Mel scale spectrogram, a visually and analytically rich representation that forms the foundational input for subsequent pattern recognition by our convolutional neural networks (CNNs). This process enhances the model’s predictive accuracy and robustness.

\subsection{Model Training and Evaluation}
\label{sec:data_preparation}

This investigation is dedicated to constructing a Convolutional Neural Network (CNN) for the regression task of predicting song popularity from audio features. Anchored in supervised learning, this study hypothesizes that patterns within audio spectrograms, when analyzed comprehensively, can serve as reliable indicators of a song's appeal. The CNN model is meticulously designed to process not just the audio data but also associated metadata, thereby enriching the prediction capability with multifaceted data insights.

\subsubsection{CNN Architecture}
The CNN architecture comprises multiple convolutional layers to effectively capture spatial hierarchies in the audio data:
\begin{itemize}
    \item \textbf{First Convolutional Layer:} This layer uses 16 filters with a kernel size of \(3 \times 3\), stride of 1, and padding to maintain the dimensionality, followed by a ReLU activation function to introduce non-linearity into the model.
    \item \textbf{Pooling Layer:} Follows each convolutional block to reduce dimensionality and enhance the extraction of dominant features, using a \(2 \times 2\) window with a stride of 2.
    \item \textbf{Subsequent Convolutional Layers:} Increase in complexity and depth, using 32 and 64 filters, respectively, to capture more detailed features from the audio spectrograms, each followed by ReLU for non-linearity.
    \item \textbf{Final Convolutional Layer:} Employs 128 filters, capturing the most abstract and representative features from the input data.
\end{itemize}

\subsubsection{Metadata Handling}
Post convolutional processing, the audio features are flattened and then combined with metadata which is preprocessed through its own dedicated layers. This integration is achieved by:
\begin{itemize}
    \item \textbf{Metadata Preprocessing Layers:} The metadata is first processed through a series of fully connected layers that transform the raw metadata into a vector of dimensions suitable for merging with audio-derived features.
    \item \textbf{Feature Concatenation:} The metadata vector is concatenated to the flattened output of the convolutional layers, allowing the network to simultaneously learn from both audio and contextual data, thus enriching the model's ability to predict song popularity based on a comprehensive set of features.
\end{itemize}

\subsubsection{Feature Processing and Prediction}
This combined feature set from both audio and metadata processing is then fed into additional fully connected layers to predict the song's popularity:
\begin{itemize}
    \item \textbf{Final Layers:} A set of fully connected layers further processes the combined features to produce a single output value representing the song's predicted popularity.
\end{itemize}

\subsubsection{Training and Validation}
To facilitate effective model training, the dataset was divided into an 80\% training set and a 20\% testing set. This division was strategic, ensuring the model was exposed to and learned from a diverse array of audio characteristics, which is critical for the nuanced understanding required for accurate popularity prediction.

An integral part of the training process was the implementation of an early stopping mechanism. This was designed to counteract overfitting by ceasing training when the validation loss ceased to decrease, thereby preserving the model’s ability to generalize to unseen data. This approach was pivotal in striking a balance between learning from the training data and maintaining adaptability to new information.

The effectiveness of the CNN model was demonstrated through its empirical performance, achieving an impressive accuracy of 95.68\% for the specified threshold. Moreover, the model's precision in estimating song popularity was further evidenced by a Mean Absolute Error (MAE) of 9.4958. This metric, in particular, offers insight into the average magnitude of errors in the predictions, underscoring the model’s reliability in practical applications.

Through the focused application of a CNN for the regression analysis of song popularity, this study not only validates the hypothesis regarding the predictive power of audio spectrograms but also showcases the model's advanced capability in integrating complex data types for insightful predictions. The achievement of high accuracy and the detailed error analysis exemplify the model’s potential in revolutionizing music analytics, illustrating the transformative impact of machine learning in predictive modeling within the music industry.

\subsection{Early Stopping}
Early Stopping is employed in our Convolutional Neural Network (CNN) to halt training when there is no significant improvement in validation loss, effectively preventing the model from overfitting. This regularization technique monitors the validation loss, \(L_v\), and terminates training based on the following condition:

\begin{equation}
 \forall e_i \in \{e_{n-p+1}, \ldots, e_n\}, L_{v}(e_i) > L_{v_{best}} - \delta
\end{equation}

Here, \(e_n\) is the current epoch, \(p\) is set to 7 epochs indicating the patience, and \(\delta\) is set to zero, implying that any non-negative change in validation loss is considered insufficient improvement. This strict criterion is set to ensure that every bit of potential overfitting is curtailed, optimizing the balance between learning from the training data and maintaining the ability to generalize to new data.

The application of Early Stopping has proven crucial for our model, which handles a complex dataset including both audio and metadata features. It helps in minimizing computational resources by avoiding unnecessary training iterations and ensures the model does not memorize the noise within the training data. Through this approach, our model achieves a robust performance, demonstrating high accuracy and generalization capability in predicting song popularity. 

Early Stopping halts training at the ideal moment to enhance model integrity and responsiveness, ensuring efficient development by preventing overtraining and aligning with performance gains.

\subsection{Model Parameters}
\label{sec:model_parameters}

The parameters for the Convolutional Neural Network (CNN) used in our study were determined through a combination of industry standards and empirical trial and error. Parameters such as batch size, learning rate, and the number of epochs were optimized based on experimental results, while the sample rate adheres to widely recognized standards for high-quality audio processing. The epochs were specifically optimized using an early stopping mechanism, which terminates training when validation losses cease to improve, preventing overfitting and ensuring the model's generalizability.

\begin{itemize}
    \item \textbf{Batch Size}: 32
    \item \textbf{Epochs}: 25
    \item \textbf{Learning Rate}: 0.001
    \item \textbf{Sample Rate}: 44100 Hz
    \item \textbf{Number of Samples}: 1967
\end{itemize}

This methodical selection process ensures that each parameter contributes optimally to the performance of the CNN, enabling robust predictions and high accuracy in estimating song popularity.

\section{RESULTS}

\subsection{Validation Error Over Epochs}

\begin{figure}[htbp]
\centering
\includegraphics[width=0.5\linewidth]{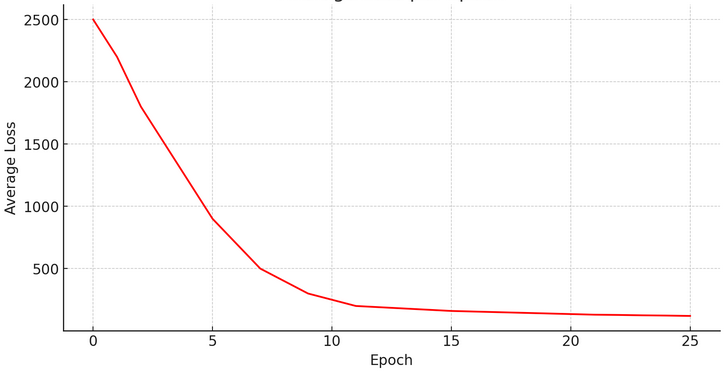}
\caption{The validation error plotted across training epochs, illustrating the model's improvement and convergence over time.}
\label{fig:val_error}
\end{figure}

The graph in Figure \ref{fig:val_error} showcases the decrease in validation error as the number of epochs increases. Initially, the error drops sharply, indicating rapid learning by the model. After around 5 epochs, the descent in error slows and begins to plateau by epoch 10, suggesting the model is approaching its convergence point. Beyond epoch 15, the error reduction is minimal, implying that the model has likely reached its optimal performance on the validation dataset. This behavior is typical in training neural networks, where initial improvements are significant and diminish over time as the model fine-tunes its parameters.

\subsection{Mean Absolute Error (MAE) Over Epochs}

\begin{figure}[htbp]
\centering
\includegraphics[width=0.5\linewidth]{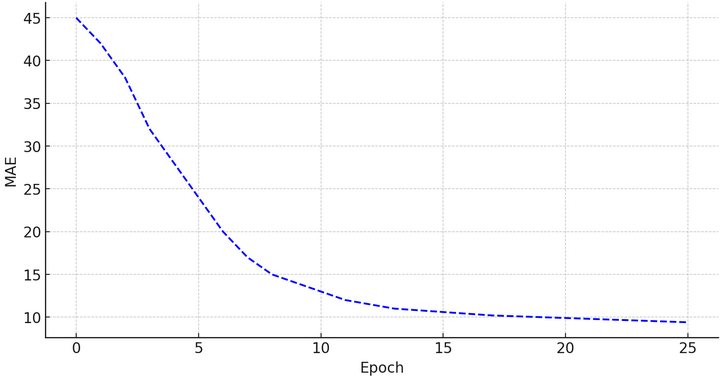}
\caption{Evolution of Mean Absolute Error (MAE) throughout the training epochs.}
\label{fig:mae_epochs}
\end{figure}

Figure \ref{fig:mae_epochs} shows the Mean Absolute Error (MAE) during the training epochs. The MAE declines sharply before stabilizing and plateauing after the 10th epoch, indicating that the model's predictive accuracy improves quickly and then gradually refines. This leveling off is indicative of the model approaching its optimal performance.

\section{Analysis of Music Popularity Predictions}

\subsection{Bar Chart for Comparing Predicted and Actual Popularity}
\begin{figure}[htbp]
\centering
\includegraphics[width=0.5\linewidth]{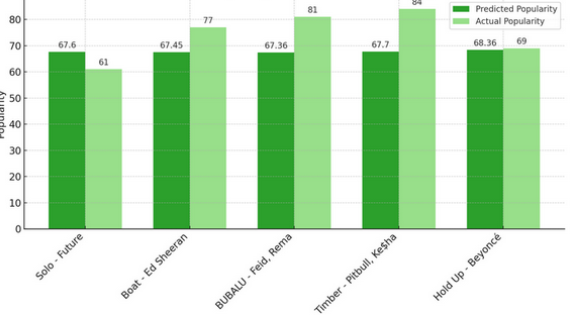}
\caption{A bar chart comparison of predicted versus actual popularity scores, highlighting the model's performance and accuracy in predictions.}
\label{fig:bar_compare}
\end{figure}

Figure \ref{fig:bar_compare} presents a bar chart that compares the predicted and actual popularity scores for various music tracks. The green bars represent the model's predictions, while the blue bars show the actual popularity scores as recorded. The numerical values atop each bar indicate the precise score. This visual representation allows for an easy comparison between the model's predictions and the true outcomes. It is evident from the chart that while some predictions are quite close to the actual figures, certain predictions deviate more noticeably. The closeness of the green and blue bars for each track indicates the level of accuracy of the model's predictive ability.

\subsection{Histogram of Prediction Errors}
\begin{figure}[htbp]
\centering
\includegraphics[width=0.9\linewidth]{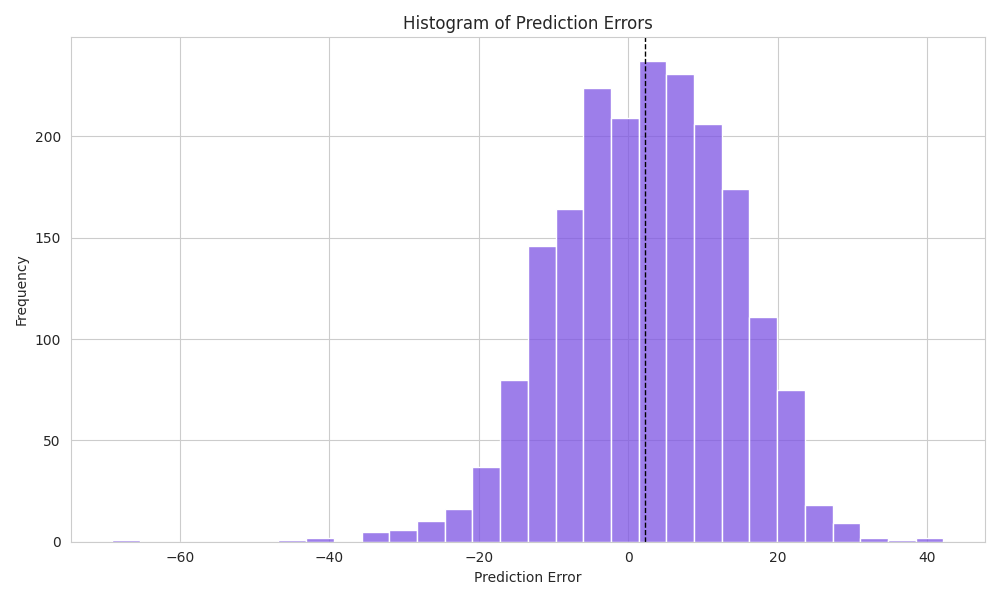}
\caption{Histogram of Prediction Errors, indicating the frequency distribution of errors in a prediction model. The vertical dashed line represents the mean error.}
\label{fig:histogram_prediction_errors}
\end{figure}

Figure \ref{fig:histogram_prediction_errors} showcases a histogram of prediction errors from a model forecasting music popularity. The histogram displays the frequency distribution of prediction errors, where each bar represents a range of error values. The vertical dashed line in the middle of the histogram marks the mean error. This chart helps in understanding the variability and typical error magnitude, which can further inform model adjustments for better accuracy.

\subsection{Heatmap of Average Popularity by Release Month and Year}
\begin{figure}[htbp]
\centering
\includegraphics[width=\linewidth]{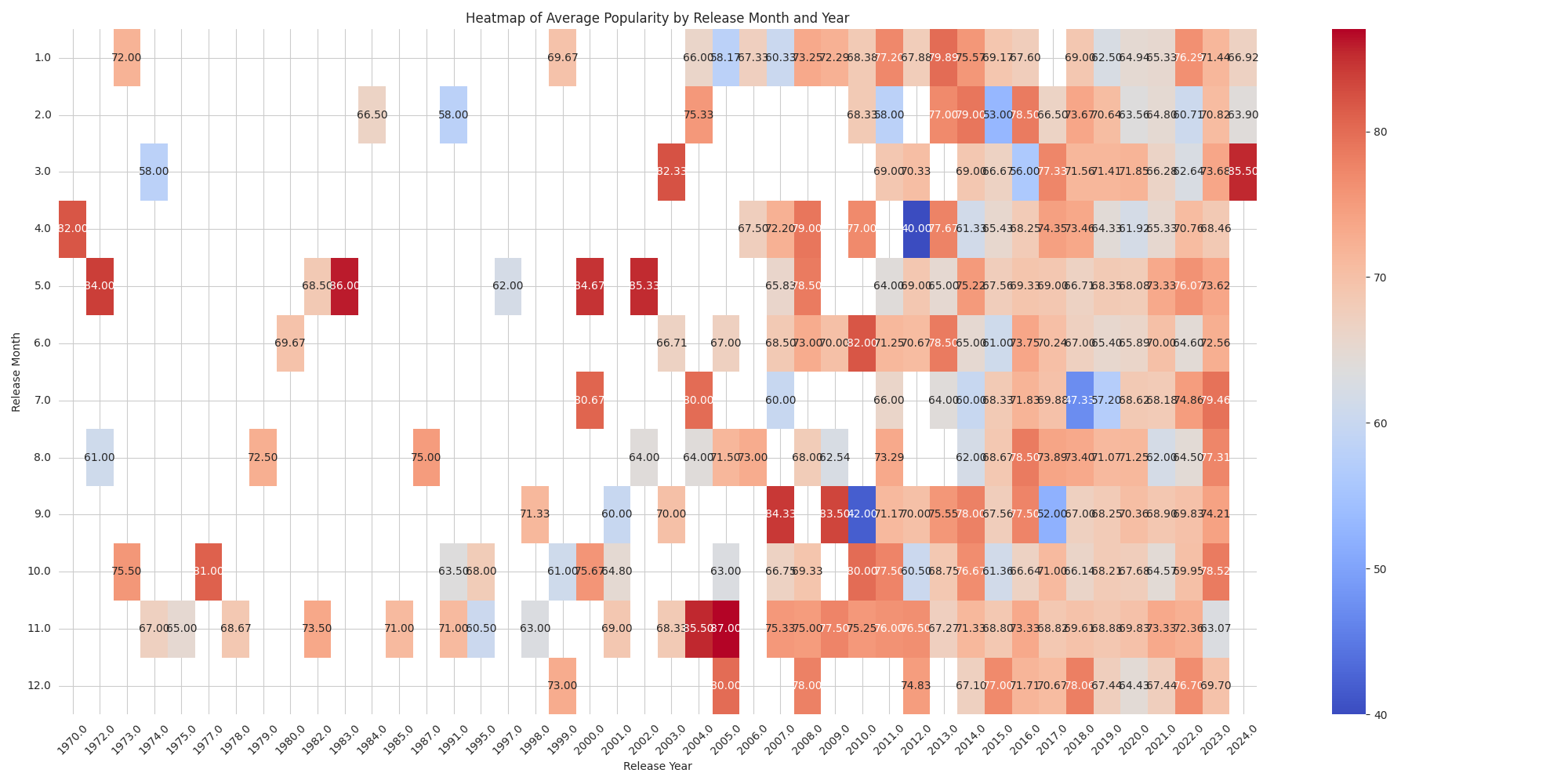}
\caption{This heatmap provides an overview of the average popularity trends for songs released in each month across different years. The color gradient represents the average popularity scores, with warmer colors indicating higher popularity and cooler colors indicating lower popularity.}
\label{fig:popularity_heatmap}
\end{figure}

Figure \ref{fig:popularity_heatmap} provides a heatmap that illustrates the average popularity trends for songs released in each month across different years. The colors on the heatmap range from blue (indicating lower popularity) to red (indicating higher popularity), with the specific popularity values noted on each block. This visual tool effectively highlights temporal trends in song popularity, allowing for analysis of seasonal effects or shifts in listener preferences over time.

\section{Analysis of Feature Correlation with Song Popularity}

Recent findings indicate that features associated with the artist, namely \textit{Artist Popularity} and \textit{Artist Followers}, have stronger correlations with song popularity than traditional musical features. Studies such as those by Kamal et al. \cite{kamal2021} and Kaneria et al. \cite{kaneria2021} have similarly identified the significant impact of artist-related metrics on song performance. As illustrated in Table~\ref{tab:correlation_popularity}, \textit{Artist Popularity} exhibits a significantly higher correlation (0.45320) with song popularity, suggesting that the artist's public profile may impact song success more than previously recognized. These insights call for a strategic shift towards enhancing artist visibility alongside musical production to optimize song performance in the market.

\begin{table}[H]
\centering
\caption{Correlation of Music Features with Popularity}
\label{tab:correlation_popularity}
\begin{tabular}{cc}
\hline
\textbf{Feature} & \textbf{Correlation with Popularity} \\ \hline
Instrumentalness & -0.03715 \\ \hline
Liveness & -0.02719 \\ \hline
Time Signature & -0.01655 \\ \hline
Acousticness & -0.01619 \\ \hline
Tempo & -0.01130 \\ \hline
Release Year & 0.00357 \\ \hline
Valence & 0.00765 \\ \hline
Energy & 0.01161 \\ \hline
Speechiness & 0.03005 \\ \hline
Loudness & 0.03983 \\ \hline
Duration (ms) & 0.08711 \\ \hline
Danceability & 0.09569 \\ \hline
Artist Followers & 0.24072 \\ \hline
Artist Popularity & 0.45320 \\ \hline
\end{tabular}
\end{table}

\section{Model Comparison}
The performance of various classification models, compared with our model, is presented in Table \ref{tab:model_metrics_detailed}. This table shows that our model has overcome the previous methods in all criteria and surpasses previous models in both precision and reliability for predicting song popularity, achieving an F1 Score of 0.9779. This suggests that enhancements in data processing and algorithmic design can substantially enhance performance, highlighting its potential for further research and practical applications in the field.

\begin{table}[h]
\centering
\caption{Detailed Metrics for Music Popularity Prediction Models Across Studies}
\label{tab:model_metrics_detailed}
\begin{tabular}{p{6.8cm}cccc}
\hline
\textbf{Model (Ref)} & \textbf{Precision} & \textbf{Recall} & \textbf{F1} & \textbf{Accuracy} \\
\hline
\textbf{Our Model} & \textbf{0.9587} & \textbf{0.9979} & \textbf{0.9779} & \textbf{0.9568} \\
SVM (Lin) (\cite{pham2016predicting}) & 0.500 & 0.706 & 0.585 & 0.762 \\
SVM (RBF) (\cite{pham2016predicting}) & 0.495 & 0.750 & 0.597 & 0.759 \\
Logistic Regression \cite{pham2016predicting}& 0.567 & 0.500 & 0.531 & 0.790 \\
Logistic Regression \cite{gao2021catching}&0.787&0.853&0.818&0.809\\
LDA (\cite{pham2016predicting}) & 0.610 & 0.529 & 0.567 & 0.807 \\
QDA (\cite{pham2016predicting}) & 0.359 & 0.618 & 0.454 & 0.647 \\
MLP (\cite{pham2016predicting}) & 0.588 & 0.441 & 0.504 & 0.793 \\
MLP \cite{gao2021catching}&0.828&0.840&0.834&0.831\\
RF \cite{saragih2023predicting} &0.69&0.70&0.69&0.7\\
RF \cite{kamal2021} & 0.83 & 0.70 & 0.76 & - \\
RF \cite{yee2022predicting}&74.5&73.2&73.1&79.6\\
RF \cite{pareek2022predicting} & 0.83 & 0.70 & 0.76 & - \\
KNN \cite{kamal2021} & 0.50 & 0.45 & 0.47 & - \\
KNN \cite{pareek2022predicting} & 0.50 & 0.45 & 0.47 & - \\
LSVC \cite{kamal2021} & 0.46 & 0.43 & 0.45 & - \\
LSVC \cite{pareek2022predicting} & 0.46 & 0.43 & 0.45 & - \\
Boosting Tree (PCA) \cite{gao2021catching}&0.817&0.847&0.832&0.828\\
Deep autoencoder+ MusicPopNet \cite{martin2020multimodal}&83.04&83.01&83.02&-\\
\hline
\end{tabular}
\end{table}

\section{Conclusion and Future Works}

In conclusion, our research leverages advanced neural processing techniques to predict music popularity based on a comprehensive analysis of audio features. By integrating social media variables and metadata from Spotify, we have developed a regression neural network model that achieves an impressive accuracy of 95.68\% and a Mean Absolute Error (MAE) of 9.4958. This model represents a significant advancement in the field of Hit Song Science, offering a novel approach that combines audio features and social media data for predicting hit songs. Moving forward, our study sets the stage for further exploration and refinement of predictive models in the music industry, with the potential to revolutionize how we understand and predict musical success in the digital age.

Future enhancements to song popularity prediction models are essential for progress in the field. Key recommendations include broadening the training dataset to capture a wider spectrum of musical features, including a varied selection of artists to introduce greater data heterogeneity, and integrating diverse data qualities to reinforce the model's adaptability. Additionally, fine-tuning model hyperparameters for optimal performance, streamlining data acquisition to ensure quality and consistency, and advancing the architectural design of the model to harness complex patterns are crucial steps. Moreover, refining training methodologies to improve learning efficacy is essential.

Implementing these strategies could significantly elevate the accuracy and reliability of future predictive models in music popularity assessment.

\bibliography{sn-bibliography}



\end{document}